\documentclass[aps,prl,amsmath,amssymb,floatfix,twocolumn, amsmath, superscriptaddress, twocolumn]{revtex4}
\usepackage{multirow}
\usepackage{bbold}
\usepackage{subfigure}
\usepackage{color}
\usepackage{mathrsfs}
\usepackage{hyperref}

\usepackage{amsfonts, relsize, color}
\usepackage{graphics}
\usepackage{graphicx}
\usepackage{hyperref}
\usepackage{color}

\newcommand{\bs}[1]{\boldsymbol{#1}}

\begin{document}
\title{Unifying Interacting Nodal Semimetals: A New Route to Strong Coupling}

\author{Shouvik Sur}
\affiliation{Department of Physics \& Astronomy, Northwestern University, Evanston, IL 60208, USA}
\affiliation{National High Magnetic Field Laboratory and Department of Physics, Florida State University, Tallahassee, Florida 32306, USA}

\author{Bitan Roy}
\affiliation{Max-Planck-Institut f\"{u}r Physik komplexer Systeme, N\"{o}thnitzer Str. 38, 01187 Dresden, Germany}
\affiliation{Department of Physics, Lehigh University, Bethlehem, Pennsylvania, 18015, USA}

\date{\today}

\begin{abstract}
We propose a general framework for constructing a large set of nodal-point semimetals by tuning the number of linearly ($d_L$) and (at most) quadratically ($d_Q$) dispersing directions. By virtue of such a unifying scheme, we identify a new perturbative route to access various strongly interacting non-Dirac semimetals with $d_Q>0$. As a demonstrative example, we relate a two dimensional anisotropic semimetal with $d_L=d_Q=1$, describing the topological transition between a Dirac semimetal and a normal insulator, and its three dimensional counterparts with $d_L=1$, $d_Q=2$. We address the quantum critical phenomena and emergence of non-Fermi liquid states with unusual dynamical structures within the framework of an $\epsilon$ expansion, where $\epsilon=2-d_Q$, when these systems reside at the brink of charge- or spin-density-wave orderings, or an $s$-wave pairing. Our results can be germane to two-dimensional uniaxially strained optical honeymcomb lattice, $\alpha$-(BEDT-TTF)$_2\text{I}_3$.   
\end{abstract}

\maketitle

\begin{figure}[t!]
\includegraphics[width=0.9\columnwidth]{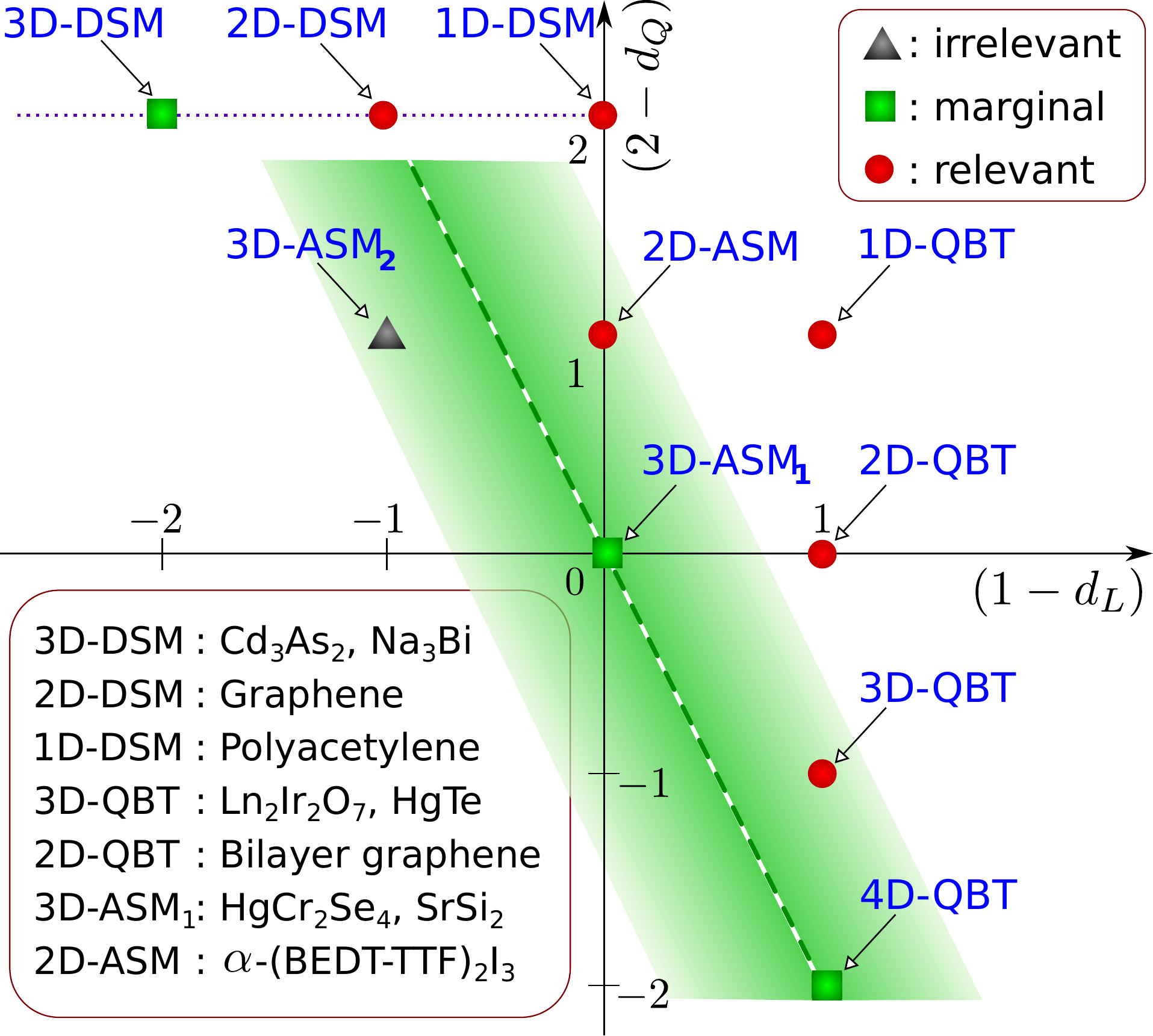}
\caption{Ten classes of nodal-point semimetals, distinguished by the number of linearly ($d_L$) and quadratically ($d_Q$) dispersing  directions. DSM, ASM and QBT denote Dirac, anisotropic and quadratic band touching semimetals, respectively. The Yukawa coupling is marginal on the dashed line, separating the fixed points where it is relevant and irrelevant. Around this line (shaded region) the quantum fluctuations engendered by the Yukawa vertex are weak, and can be controlled by an $\epsilon$ expansion. The $d_Q=0$ line is special due to the presence of full spacetime Lorentz symmetry~\cite{Comment:ConnectionDSM}. Lower inset: Material realizations of various semimetals.
}~\label{fig:theory-space}
\end{figure}

\emph{Introduction:} Over the last decade, semimetals with a discrete number of band touching or nodal points have become increasingly important for the understanding of topological phases of matter~\cite{Kane-RMP, QiZhang-RMP, Armitage-RMP, volovik-book}. In this paradigm,  Dirac semimetals (DSMs) play an important role~\cite{Shen:Dirac-book}. Semimetals are, however, more general states of matter~\cite{bernevig:NewFermions}, and the bands in the neighborhood of the nodes may develop finite curvatures along one or more directions, see Fig.~\ref{fig:theory-space}.

\begin{figure}[t!]
\includegraphics[width=0.95\columnwidth]{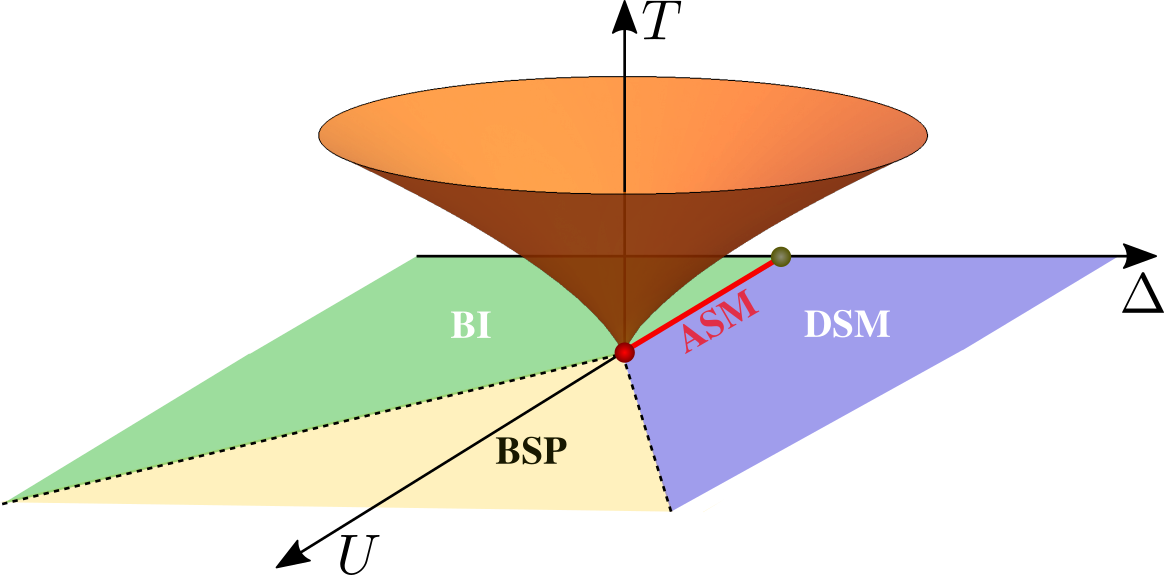}
\caption{ A schematic finite temperature ($T$) phase diagram diagram of a two dimensional interacting ASM (red line), separating a DSM and a band insulator (BI). In a noninteracting system ($U=0$) the transition between these two phases takes place through a quantum critical point located at $\Delta=0$ (gray circle). For sufficiently strong interaction ($U$) a broken symmetry phase (BSP) sets in through a multicritical point (red circle), where DSM, BI, ASM and a BSP meet. Universality classes of various transitions across this fixed point are summarized in Table~\ref{tab:exponents}. The critical cone accommodates a quasiparticleless metallic state or NFL.  
}~\label{fig:PhaseDiagram}
\end{figure}

A positive band-curvature enhances the density of states (DOS), which in turn strengthens the effects of Coulomb interactions. Therefore, semimetallic systems, falling outside the realm of linearly dispersing Dirac fermions, constitute a rich ground for exotic emergent phenomena, which we unveil (mainly) for an interacting two-dimensional anisotropic semimetal (ASM), displaying both linear and quadratic band dispersions, see Fig.~\ref{fig:PhaseDiagram}. Our results can be relevant for uniaxially strained optical honeycomb lattice~\cite{opticallattice-1, opticallattice-2, opticallattice-3, Montambaux-blochzener}, $\alpha$-(BEDT-TTF)$_2\text{I}_3$~\cite{organic-1}, black phosphorus~\cite{blackphosphorus-1} and at the TiO$_2$/VO$_2$ interface~\cite{oxide-1, oxide-2}, subject to strong Hubbardlike interactions~\cite{roy-foster-PRX}. Throughout we neglect the long-range tail of the Coulomb interaction~\cite{isobe2016, moon2016}, which is justified for ultracold neutral fermionic atoms in optical lattices or in the presence of a screening or proximate gate.

Traditionally the spatial dimensionality is tuned to gain insights into the strongly coupled phases of matter~\cite{bollini1972,veltman1972,wilson1972}. For example, strong interactions between fermionic and bosonic (such as, collective modes~\cite{Footnote:Collectivemode}), degrees of freedom can be addressed in terms of the deviation from a three-dimensional DSM (3D-DSM)~\cite{footnote:3d-DSM, zinn-justin} along the dotted line in Fig.~\ref{fig:theory-space}. Here we formulate a distinct theoretical approach that ties together a wide class of interacting nodal semimetals by continuously varying the number of quadratically ($d_Q$) and linearly ($d_L$) dispersing directions, see Fig.~\ref{fig:theory-space}. This classification scheme allows us to identify (1) a three-dimensional ASM with $d_L=1$ and $d_Q=2$ (3D-ASM$_1$), and (2) a four-dimensional quadratic band touching or Luttinger semimetal (4D-QBT) with $d_L=0$ and $d_Q=4$, as two fixed points where  boson-fermion interactions are dimensionless or marginal. Moreover, we find an entire line of such marginal fixed points, the `marginal line' (dashed line Fig.~\ref{fig:theory-space}), about which perturbative $\epsilon$ expansions can be performed by tuning $d_L$ and/or $d_Q$ to gain controlled access (within the green shaded region) to all non-Dirac semimetals with at most quadratic dispersion, coupled to gapless bosonic degrees of freedom. As a demonstrative example, we here address the quantum criticality for a strongly interacting two-dimensional ASM (with $d_L=d_Q=1$) by performing an $\epsilon$ expansion about 3D-ASM$_1$, where $\epsilon=2-d_Q$. Such an $\epsilon$ expansion is distinct from the one tailored for Lorentz symmetric Dirac systems~\cite{zinn-justin}. Our results are summarized in Table~\ref{tab:exponents}. A schematic phase diagram is shown in Fig.~\ref{fig:PhaseDiagram}.

\emph{Model:} The effective single particle Hamiltonian describing a generic, $d$-dimensional, non-interacting nodal-point semimetal
is given by
\begin{equation}~\label{eq:h0-1}
H_0(k_i) = \sum_{l=1}^{d_L} \Gamma_l k_l + \sum_{n=1}^{n_{\rm max}} \Gamma_{d_L +n} \; \varepsilon_n( k_{d_L +1}, \ldots, k_d),
\end{equation}
where $k_i$ are components of momentum (measured from the band touching point), $\{\Gamma_i \}$ is a set of mutually anticommuting matrices, $\varepsilon_n$ are quadratic functions of their arguments, $d= d_L + d_Q$ and $n_{\rm max} \leq d_Q$. Short-ranged or Hubbardlike interactions can trigger a symmetry breaking phase by generating a mass term $\sum_{a=1}^{N_b} M_a \Upsilon_a$, where $\{\Upsilon_a \}$ is another set of mutually anticommuting matrices  satisfying  $\{\Upsilon_a, \Gamma_i \} = 0$. $N_b$ is the number of order parameter components. Onset of such an ordering leads to a fully gapped state and the magnitude of $\vec{M}$ determines condensation energy gain. Here we focus on the quantum critical point (QCP) controlling such a transition.

Near a QCP, relevant degrees of freedom are gapless nodal fermions and bosonic order parameter fluctuations. Since order parameters are composite objects of fermions, these two degrees of freedom are coupled via Yukawa interaction ($g_{_0}$). Therefore, the dynamics at intermediate energies is controlled by the effective action 
\begin{eqnarray}~\label{eq:S-d}
\tilde{S} &=& \sum_{n=1}^{N_f} \int_k \psi_{n,k}^{\dagger} G^{-1}_0(k) \psi_{n,k} +  \frac{1}{2} \sum_{a=1}^{N_b} \int_q  D_0^{-1}(q) \phi_{-q}^{a} \phi_{q}^{a}  \nonumber \\
&+& \frac{g_{_0}}{\sqrt{N_f}} \sum_{n=1}^{N_f} \sum_{a=1}^{N_b} \int_{k,q} \: \phi_q^a  \left( \psi_{n,k+q}^{\dag}   \Upsilon_a \; \psi_{n,k} \right) + \ldots,
\end{eqnarray}
where $k \equiv (k_0, k_1, \ldots, k_d)$, with $k_0$ being the Euclidean frequency. Here, $\psi_{n,k}$ ($\vec \phi_q$) is a  Grassman (real vector) field that represents the $n$th copy of the fermionic modes (order parameter). The ellipses represent higher order terms (including $u |\vec \phi|^4$) that are unimportant for the present analysis. We have also introduced $N_f$ identical copies of fermions. The bare fermionic and bosonic propagators are $G_0(k)= \left[ i k_0 + H_0(k_1, \ldots, k_d)\right]^{-1}$ and 
\begin{equation}
D_0(q) = \left[ c^2 \left( q_0^2 + \sum_{l=1}^{d_L} q_{l}^2 \right) + \sum_{n=d_L + 1}^{d_Q} q_{n}^2 + |\vec M|^2 \right]^{-1}, \nonumber 
\end{equation}
respectively. The parameter $c$ encodes an anisotropy of the bosonic dynamics, stemming from the anisotropic fermionic dispersion (when $d_Q > 0$).

\emph{Scaling:} We  define a $(d_L+1)$-dimensional frequency-momentum vector ${\bs{k}} \equiv (k_0, k_1, \ldots, k_{d_L})$ and a $d_Q$-dimensional momentum ${\bs{K}} \equiv (k_{d_L + 1}, \ldots, k_d)$. The noninteracting part of the action is invariant under the scaling $[{\bs{K}}] = 1$ and $[{\bs{k}}]= z_{\bs{k}}$, yielding  
\begin{align}~\label{Eq:Scalingdim}
[c] = 1 - z_{\bs{k}},\; [M] = 1, [g_{_0}] = 1 -  \frac{z_{\bs{k}}(d_L - 1) + d_Q}{2},
\end{align}
so that $X \mapsto X' = X e^{[X] \ell}$ under rescaling, where $\ell$ is the logarithmic length scale~\cite{footnote:[u]}. The noninteracting fixed point possesses a full spacetime Lorentz symmetry only when $d_Q = 0$  and $z_{\bs{k}} = 1$, representing DSMs for which the Yukawa coupling is marginal at $(d_L, d_Q) = (3,0)$. Away from the DSM-line, we set $z_{\bs{k}} = 2$ at the non-interacting fixed point such that the fermionic propagator remains invariant under the tree-level scaling. Therefore, the Yukawa coupling is marginal on the line, $2d_L + d_Q = 4$, which passes through $(d_L, d_Q) = (1,2)$ and $(0,4)$, respectively representing a 3D-ASM$_1$ and 4D-QBT, see Fig.~\ref{fig:theory-space}. Our proposed formalism is applicable for relativistic DSMs, upon setting $d_Q=0$ from outset.

\emph{$\epsilon$ expansion:} We now focus on the QCP, residing on the critical hyperplane $|{\vec M}|^2 =0$. Since the DSM-line is relatively well understood~\cite{zinn-justin}, we focus on the $d_Q > 0$ region. The deviations from the `marginal line' ($[g_{_0}]=0$) along $d_L$ and $d_Q$ directions can be, respectively, parameterized by $\epsilon_L=\bar d_L - d_L$ and $\epsilon_Q = \bar d_Q-d_Q$, where $2 \bar d_L + \bar d_Q = 4$. Various interacting fixed points, where the Yukawa coupling is dimensionful, can be reached by tuning $\epsilon_L$ and $\epsilon_Q$ independently. At a formal level, it is accomplished through dimensional regularization, and the RG equations can be derived within the minimal subtraction scheme.~This methodology (for additional details see Supplemental Materials~\cite{supplementary}) differs substantially from the ones applied to Lorentz symmetric DSMs~\cite{zinn-justin}, while bearing similarities to those for critical Fermi surfaces~\cite{Dalidovich2013,sur-Lee-2014}.

\begin{table}[t!]
\begin{tabular}{|c || c | c | c |}
\hline 
 & $N_b=1$ (Ising)  &  $N_b=2$ (XY)  &  $N_b=3$ (Heisenberg)   \\ 
\hline \hline
$g_*^2$ & $ f_1(N_f)~ \epsilon^{1+1/N_f}$ & $ (1- \frac{1}{N_f})~ \epsilon$ & $  f_3(N_f) ~\epsilon^{1-1/N_f}$ \\ 
\hline 
$z_{\bf k}$~ & ~$2 +   \frac{1}{2N_f} ~\epsilon  \ln{\epsilon}$ & ~$2+   \frac{1}{N_f} ~\epsilon  \ln{\epsilon}$ ~ &   ~$2 + \frac{3}{2N_f} ~\epsilon  \ln{\epsilon}$   \\
\hline
~$\eta_\psi$~ &  ~$-\frac{3}{4N_f} ~\epsilon  \ln{\epsilon}$~ &   ~$-\frac{1}{2N_f} ~\epsilon  \ln{\epsilon}$~ & ~$-\frac{9}{4N_f} ~\epsilon  \ln{\epsilon}$ \\
\hline
~$\eta_\phi$~ & ~  ~$-\frac{1}{2N_f} ~\epsilon  \ln{\epsilon}$~  & $-\frac{1}{N_f} ~\epsilon  \ln{\epsilon}$ & ~$-\frac{3}{2N_f} ~\epsilon  \ln{\epsilon}$ \\
\hline
$\nu_M^{-1}$ & ~$2- \epsilon $ & $2-\epsilon$ & $2-\epsilon $  \\
\hline
$\nu_\Delta^{-1}$ & $2+\frac{1}{N_f} ~\epsilon  \ln{\epsilon}$ & $2+\frac{2}{N_f} ~\epsilon  \ln{\epsilon}$  &   $2+\frac{3}{N_f} ~\epsilon  \ln{\epsilon}$ \\
\hline \hline
\end{tabular}
\caption{Fixed points ($g^2_\ast$), anomalous dimensions ($z_{\bs k}, \eta_\psi, \eta_\phi$), and correlation length exponents ($\nu_M,\nu_\Delta$), obtained in the limit $\epsilon \ll 1, N_f \gg 1$, where $\epsilon=2-d_Q$ and $N_f$ is fermion flavor number~\cite{Table-exponents}. Respectively $N_b=1, 2$ and $3$ are pertinent near charge-density-wave, superconducting $s$-wave and spin-density-wave orderings. While $z_{\bs k}$ controls band curvature, $\eta_\psi (\eta_\phi)$ is fermionic (bosonic) anomalous dimension. The correlation length exponents for ASM-BSP (DSM-BI) quantum phase transitions are $\nu_M$ ($\nu_\Delta$), see Fig.~\ref{fig:PhaseDiagram}. Here $f_1(x)=2 \pi^2( \pi^2 e^{1/2}/8)^{1/x}$ and $f_3(x) = 2 \pi^2(\pi^2 e^{5/2}/8 )^{-1/x}$.
}~\label{tab:exponents}
\end{table}

\emph{Interacting ASM}: To exemplify the framework discussed so far, we focus on 3D-ASM$_1$, describing either (1) the topological QCP separating a nodal-line semimetal and a band insulator (BI)~\cite{Fu-NLSM, Roy-NLSM, Sur-NLSM} or (2) double Weyl semimetal~\cite{DWSM-1,DWSM-2,DWSM-3}. We focus on the former system, for which the effective single-particle Hamiltonian reads 
\begin{align}~\label{eq:h0-2}
H_0(k_x, k_y, k_z) =  \sigma_0 \: \left[ k_x \tau_1 + (k_y^2 + k_z^2 - \Delta) \tau_2 \right].
\end{align}
Two sets of Pauli matrices $\{ \sigma_\mu\}$ and $\{ \tau_{\mu} \}$ respectively operate on the spin and sublattice or orbital degrees of freedom, where $\mu=0, \cdots, 3$. Respectively, for $\Delta>0$ and $\Delta<0$, the system describes a nodal-loop semimetal of radius $\sqrt{\Delta}$ and a BI. The transition between these two phases takes place at $\Delta=0$ and described by 3D-ASM$_1$. In the absence of $k_z$ (i.e., in 2D), the above single-particle Hamiltonian captures the transition between a two-dimensional DSM and a BI~\cite{2D-ASM-1}. At the topological QCP, $\Delta$ is a relevant perturbation and it plays the role of a tuning parameter. Therefore, emergent quantum critical phenomena in a strongly interacting 2D-ASM with $d_L=d_Q=1$ can be addressed by an $\epsilon$ expansion about its three dimensional counterpart (3D-ASM$_1$), with $\epsilon=2-d_Q$. Hence, this approach is distinct from the one employed for Dirac systems ($d_Q=0$), where dimensionality of the system is continuously tuned with $\epsilon=3-d_L$~\cite{zinn-justin}. For $\Delta>0$, the system describes a Lorentz symmetric ($z_{\bs{k}}=1$) Dirac (nodal-loop) semimetal at low-energies in $d=2(3)$ and an ASM ($z_{\bs{k}}=2$) at high-energies. Hence, the above model is also suitable for describing Lifshitz-type quantum field theories~\cite{Alexandre-review}.

\begin{figure}[!t]
\centering
\subfigure[]{\includegraphics[width=0.2\textwidth]{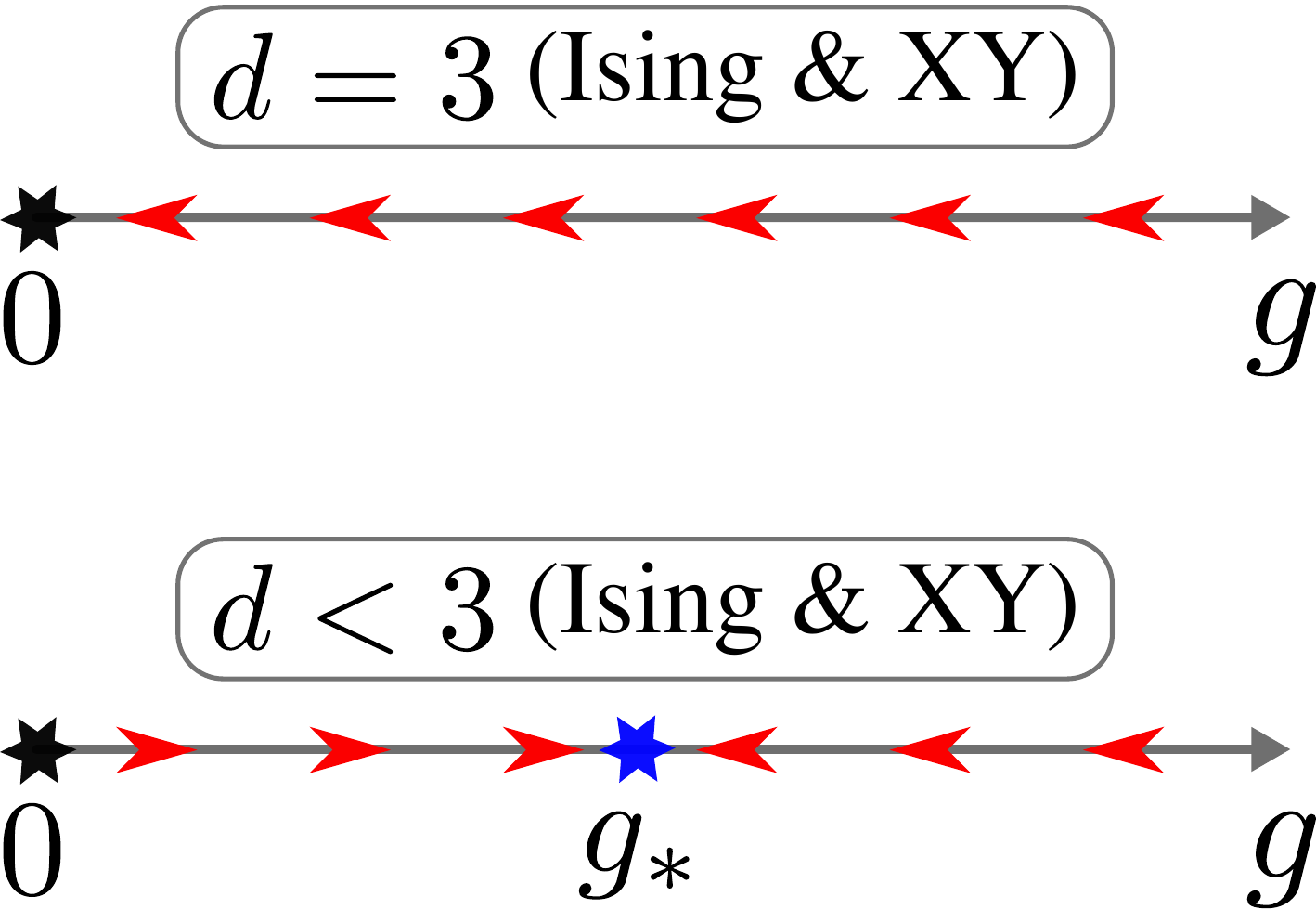}}
\hfill
\subfigure[]{\includegraphics[width=0.2\textwidth]{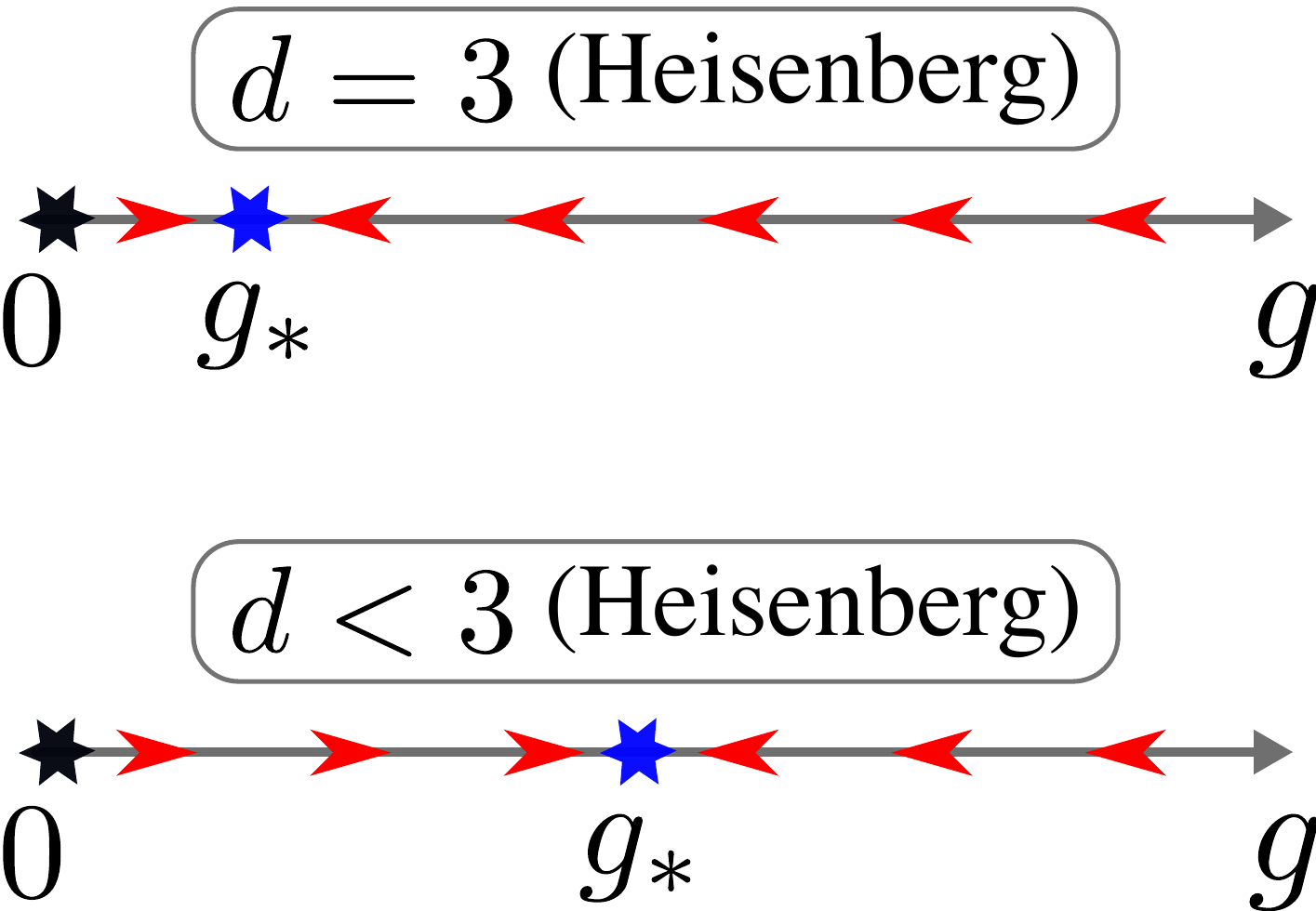}}
\caption{Schematic plot of the RG flow (red arrows) of $g(\ell)$ [see Eq.~(\ref{eq:RG-eqn})]. (a) For $N_b = 1$ (Ising) and $2$ (XY) the non-interacting fixed (black asterix) point is stable in $d=3$, while a stable quasiparticleless interacting fixed point (blue asterix) is obtained only for $d<3$. (b) By contrast, for $N_b = 3$ (Heisenberg) a stable interacting fixed point is already present in $d=3$, which shifts to a stronger coupling for $d<3$.
}~\label{fig:flow}
\end{figure}

For sufficiently strong interactions, the ASMs can terminate at a multicritical point, where a gapless topological phase (DSM or nodal-loop semimetal), BI, ASM and a broken symmetry phase meet, see Fig.~\ref{fig:PhaseDiagram}. If we set the renormalized value of $\Delta$ to be zero, the effective action at an intermediate momentum scale $\mu$ takes the form 
\begin{align}~\label{eq:S-d-min}
S &= \sum_{n=1}^{N_f} \int_k \psi^\dagger_{n,k}  G_0^{-1}(k) \psi_{n,k} 
+ \frac{1}{2} \sum_{a=1}^{N_b} \int_q |{\bs{Q}}|^2 \: \phi_{-q}^a \phi_q^a   \nonumber \\
& + \frac{g \mu^{(2-d_Q)/2}}{\sqrt{N_f}} \sum_{n=1}^{N_f} \sum^{N_b}_{a=1}\int_{k,q} \phi^a_q \: \left( \psi^\dagger_{n,k+q} \Upsilon_a \psi_{n,k} \right),
\end{align}
where $-\ln \mu \equiv \ell$ and $g \equiv g_{_0} \mu^{(d_Q-2)/2}$ is the dimensionless Yukawa coupling. Anticipating the correct $\boldsymbol{q}$ dependence of the boson's dynamics generated by quantum fluctuations, we set the irrelevant parameter $c=0$ at the tree-level~\cite{supplementary}. In the close proximity to charge- and spin-density wave orderings $N_b=1$ and $3$, and ${\boldsymbol \Upsilon}=\sigma_0 \tau_3$ and ${\boldsymbol \sigma} \tau_3$, respectively. With an appropriate redefinition of the spinor basis the above effective theory is applicable near an $s$-wave pairing, for which $N_b=2$ and ${\boldsymbol \Upsilon}= {\boldsymbol \sigma}_{_\perp} \tau_3$, where ${\boldsymbol \sigma}_{_\perp} = \left( \sigma_1, \sigma_2 \right)$. Within the framework of an extended Hubbard model, containing onsite (repulsive and attractive) and nearest-neighbor interactions, these three phases are the most prominent ones~\cite{roy-foster-PRX, Roy-NLSM, 2DASM-1, 2dASM-3}.

The Yukawa coupling becomes progressively more relevant as $d_Q \to 1$, the effects of which are systematically incorporated by an $\epsilon$ expansion, with $\epsilon \equiv \epsilon_Q =2-d_Q$. The `small' parameter $\epsilon$ measures the deviation from the upper critical dimension $d_Q=2$, where $[g_{_0}]=0$ for $d_L=1$. In the neighborhood of the upper critical dimension, the non-analyticities resulting from the anisotropic dispersion decouple from those originating from the relevance of $g$ at the leading order, allowing a more controlled access to asymptotically low energy than in $d=2$. Moreover tuning $d_Q$ preserves the rotational symmetry in $\bs k$-space, which implies  that the dynamical critical exponent defined with respect to the linearly dispersing direction is fixed at \emph{unity}. In addition, the number of Pauli matrices necessary to define a $d$-dimensional local critical theory is independent of $d$, permitting $d$ to be truly arbitrary.

\emph{RG \& Fixed points:} Here the dominant infrared (IR) processes that cut off the putative divergences in the $c \to 0$ limit occur in the RPA channel, and the implicit resummation amounts to replacing $D_0^{-1}(q)$ by the damped bosonic propagator~\cite{supplementary} 
\begin{equation}
D^{-1}(q) = |{\bs{Q}}|^2 + g^2 |{\bs{q}}|/16. 
\end{equation}
Even though the resultant dynamics lacks point-like bosonic excitations, the fermions are still weakly coupled to the bosons. And to the linear order in $\epsilon$, we obtain the following RG flow equation for $g$
\begin{equation}~\label{eq:RG-eqn}
\frac{d g}{d \ell} = \frac{\epsilon}{2} \; g - \frac{g^3}{4\pi^2}  \left[ 1 + \frac{3N_b - 4}{2 N_f}  + \frac{N_b - 2}{N_f} \ln{\left(\frac{g^2}{16} \right) } \right].
\end{equation} 
While the vertex correction vanishes for $N_b=2$, it screens (anti-screens) the Yukawa coupling for $N_b = 1 (3)$~\cite{sur2016a}. The unusual $\ln g^2$ term  results  from the dynamical IR adjustments, and implies that a naive perturbative expansion around the non-interacting fixed point cannot access the asymptotically low energy regime~\cite{supplementary}.

At $\epsilon = 0$, the noninteracting fixed point at $g^2=0$ is stable for $N_b = 1$ and $2$, but it becomes unstable to an interacting fixed point at $g^2_\ast= 16 \; e^{-(N_f+5/2)}$ for $N_b= 3$, see Fig.~\ref{fig:flow}. This interacting fixed point is stabilized by a balance between the anomalous dimensions and vertex correction, and is inaccessible within a naive one-loop expansion. Therefore, in a three-dimensional interacting nodal-loop semimetal, residing in the close proximity to a topological QCP, only the transition to a spin density-wave phase can be non-Gaussian. By contrast, for $d_Q<2$ stable IR fixed points are obtained for $N_b=1,2,3$. Although the peturbative expansion is controlled in terms of $\epsilon$, to simplify the results, we take the large $N_f$ limit at fixed $\epsilon$ and retain the leading order terms in $1/N_f$. The results are displayed in Table~\ref{tab:exponents}. Only the superconducting or XY QCP is conventional since $g_*^2 \propto \epsilon$. By comparison, quantum fluctuations are weakened (strengthened) in the neighborhood of the itinerant charge (spin) density-wave or Ising (Heisenberg) QCP.

\emph{Non-Fermi liquid (NFL):} The leading order RG analysis yields the following nontrivial anomalous dimensions 
\begin{align}~\label{eq:anomalousdim}
z_{\bs{k}}=2+a {\bar g}^2, \:
\eta_\psi =-b {\bar g}^2, \:
\eta_\phi=-\left( a-2 \frac{N_f}{N_b} \right) {\bar g}^2,
\end{align}
where $\bar g^2 \equiv \frac{N_b}{2N_f} \frac{g^2}{(2\pi)^2}$, $a=7+2 \ln (g^2/16)$ and $b=10+3 \ln (g^2/16)$~\cite{supplementary}.  While $z_{\bs{k}}$ controls the curvature of the dispersion according to $E_\pm(k_x, {\bs{K}}) \sim \pm \sqrt{k_x^2 + |{\bs{K}}|^{2 z_{\bs{k}}}}$, the fermionic ($\eta_\psi$) and bosonic ($\eta_\phi$) anomalous dimensions, respectively, modify the corresponding two-point correlation function  according to $\langle \psi_k \psi_k^\dag \rangle \propto |{\bs{k}}|^{-1+2\eta_\psi}$ and $\langle \phi_k \phi_{-k} \rangle \propto |{\bs{k}}|^{-1+2\eta_\phi}$. Hence, the residue of the fermionic and bosonic quasiparticle pole vanishes at this fixed point, indicating emergence of a NFL in its vicinity. Since the anomalous dimensions of the fields are positive (see Table~\ref{tab:exponents}), compared to the non-interacting limit, the fluctuations are more localized in the real space at the stable IR fixed points. Next, we discuss the imprints of such quasiparticleless metallic states at finite temperature and frequency within the critical cone, see Fig.~\ref{fig:PhaseDiagram}.

The correlation lengths (along $\hat{x}$) for the topological (between DSM or nodal-loop semimetal and BI) and ordering (between ASM and a broken symmetry phase) transitions are given by $\xi_\Delta \sim |\Delta|^{-\nu_\Delta z_{\bs{k}}}$ and $\xi_M \sim |M^2|^{-\nu_M z_{\bs{k}}}$, respectively. On approaching the multicritical point they diverge faster than the mean-field expectation $\xi_X \sim |X|^{-1}$, where $X \in \{\Delta, M \}$. Since the dynamical exponent with respect to $x$ is $1$, the temperature scale ($T_\ast$) associated with the transitions, tuned by $X=\Delta$ and $M$, is given by $T_\ast \sim \xi_X^{-1}$. At the interacting fixed points, the spectral density function measured at the erstwhile band-touching point scales as ${\mathcal A}(\omega) \sim \omega^{-1 + 2\eta_\psi}$. The dynamical structure factors for the order parameter fluctuations goes as ${\mathcal S}(\omega) \sim \omega^{-1 + 2\eta_\phi}$. Within the critical cone the specific heat and compressibility scale as $C \sim T^{1+d_Q/z_{\bs{k}}}$ and $\kappa \sim T^{d_Q/z_{\bs{k}}}$, respectively. And specifically in two dimensions the interband optical conductivity of the NFL along $\hat{x}$ and $\hat{y}$ are respectively given by $\sigma_{xx} \sim \omega^{-1+ z^{-1}_{\bs{k}} + 4\eta_\psi}$ and $\sigma_{yy} \sim \omega^{1- z^{-1}_{\bs{k}} + 4\eta_\psi}$.

\emph{Discussion:} We introduced a new classification scheme to tie together a large set of nodal-point semimetals, by tuning the number of linearly ($d_L$) and quadratically ($d_Q$) dispersing directions [Fig.~\ref{fig:theory-space}]. Such a unifying scheme allows us to capture emergent quantum critical phenomena in strongly interacting semimetallic systems that lie beyond conventional DSMs. We also identify suitable $\epsilon$ expansions to access the quantum critical regimes in non-Dirac systems in a controlled fashion. As a demonstrative example, we address the scaling behavior of 2D-ASM ($d_L=d_Q=1$) and 3D-ASM$_1$ ($d_L=1, d_Q=2$), residing at the brink of spontaneous symmetry breaking, in terms of $\epsilon=2-d_Q$ which continuously interpolates between these two semimetals [see Table~\ref{fig:theory-space}]. Observed anisotropic band dispersion in black phosphorus~\cite{blackphosphorus-1}, where the strength of electronic interactions can be tuned by applying hydrostatic pressure, for example, along with tunable band structure and Hubbard interaction in optical honeycomb lattice~\cite{opticallattice-1,opticallattice-2,opticallattice-3} and pressured $\alpha$-(BEDT-TTF)$_2\text{I}_3$~\cite{organic-1, organic-2, organic-3} make predicted NFL scaling for a two-dimensional ASM observable in real materials, which can also be tested in quantum Monte Carlo simulations of extended Hubbard model~\cite{QMC-1,QMC-2,QMC-3, QMC-4}.

We close the discussion with a qualitative comparison between quantum criticalities of a Fermi surface and an ASM. In both systems, damping of bosonic dynamics is responsible for the emergent NFL behavior. In the former system, however, the damping scale competes with that of a spontaneous symmetry breaking due to a finite DOS, which generally results in superconductivity at low  temperatures~\cite{Mross2015, Kivelson2017}. By contrast, in an ASM the NFL behavior is expected to survive down to  arbitrary low energies (a `naked' QCP) due to the vanishing DoS, while the NFL scaling can be observed at temperatures $T>\tilde{\mu}$ for a finite doping ($\tilde{\mu}$)~\cite{roy-juricic-2019}. A \emph{qualitatively} similar situation arises in a Fermi liquid quantum criticality~\cite{Kivelson2017}. Moreover, the non-interacting Hamiltonian of a $d_Q$ dimensional ASM [Eq.~(\ref{eq:h0-1})] is similar to those in patch-theories of Fermi surface QCPs~\cite{Mandal2015}, with the difference arising from the pole structure of the respective fermion propagators. Nonetheless, similar quantum dynamical structures at low energies suggest intriguing parallels between these two classes of criticalities, which may provide new insights into Fermi surface criticalities.

\emph{Acknowledgments:} S.S. was supported in parts by the National Science Foundation No. DMR-1442366, and the start-up funds of Pallab Goswami from Northwestern University. S.S. thanks Sung-Sik Lee, Peter Lunts and Sedigh Ghamari for helpful suggestions.


\begin{thebibliography}{}
\bibitem{footnote:[u]} For $d_Q>0$, $[u]= 4-\{2(d_L + 1) + d_Q\}$ which makes the $|\vec{\phi}|^4$ terms irrelevant for non-Dirac semimetals.
\bibitem{Table-exponents} The interacting fixed point ($g^2_\ast$) exists for any $N_b$ and $N_f$ when $0< \epsilon \leq 1$ [see Eq.~(\ref{eq:RG-eqn})], although the  anomalous dimensions tend to vanish close to $\epsilon = 1$. This feature is, however, only an artifact of leading order $\epsilon$ and $1/N_f$ expansions. The precise values of these quantities can only be estimated after accounting for higher order corrections and subsequently performing a series resummation.
\end{thebibliography}
\end{document}